\documentclass{article}
\usepackage{amsmath,graphicx}
\usepackage{multirow}
\usepackage{INTERSPEECH2019}
\usepackage[utf8]{inputenc}
\usepackage{url,times,booktabs}
\usepackage{comment}
\usepackage{tabularx}
\usepackage{makecell}
\usepackage{cite}

\title{Short utterance compensation in speaker verification via cosine-based teacher-student learning of speaker embeddings}

\name{Jee-weon Jung$^*$\thanks{$^*$These authors contributed equally}, Hee-Soo Heo$^*$, Hye-jin Shim, and Ha-Jin Yu$^\dag$\thanks{$^\dag$ Corresponding author}\thanks{This work was supported by the Technology Innovation Program (10076583, Development of free-running speech recognition technologies for embedded robot system) funded by the Ministry of Trade, Industry \& Energy(MOTIE, Korea)}}
\address{School of Computer Science, University of Seoul, South Korea}
\email{jeewon.leo.jung@gmail.com,
zhasgone@naver.com,
shimhz6.6@gmail.com,
hjyu@uos.ac.kr}

\begin{document}
\maketitle
\begin{abstract}
The short duration of an input utterance is one of the most critical threats that degrade the performance of speaker verification systems. 
This study aimed to develop an integrated text-independent speaker verification system that inputs utterances with short duration of 2 seconds or less. 
We propose an approach using a teacher-student learning framework for this goal, applied to short utterance compensation for the first time in our knowledge. 
The core concept of the proposed system is to conduct the compensation throughout the network that extracts the speaker embedding, mainly in phonetic-level, rather than compensating via a separate system after extracting the speaker embedding. 
In the proposed architecture, phonetic-level features where each feature represents a segment of 130 ms are extracted using convolutional layers. 
A layer of gated recurrent units extracts an utterance-level feature using phonetic-level features. 
The proposed approach also adopts a new objective function for teacher-student learning that considers both Kullback-Leibler divergence of output layers and cosine distance of speaker embeddings layers. 
Experiments were conducted using deep neural networks that take raw waveforms as input, and output speaker embeddings on \texttt{VoxCeleb1} dataset. 
The proposed model could compensate approximately 65 \% of the performance degradation due to the shortened duration. 
\end{abstract}

\section{Introduction}
\label{sec:1}
Recent speaker verification systems generally work based on utterance-level features such as i-vectors, or speaker embeddings from deep neural networks (DNNs) such as x-vector system \cite{dehak2011front, DeepSpeaker, snyder2018x}. 
In utterance-level features extracted from short utterances, uncertainty exist owing to the insufficient phonetic information, which is a well-known factor of performance degradation of speaker verification systems \cite{kanagasundaram2011vector}. 
To compensate for this uncertainty caused by short utterances, Saeidi \textit{et al.} proposed a propagation method in the i-vector space \cite{uncertaintypropagation}. 
Yamamoto \textit{et al.}, proposed a DNN-based compensation system that transforms an i-vector extracted from a short utterance into an i-vector corresponding to a long utterance. 
In Yamamoto's research, it was shown that phonetic information can be effectively used for compensating short utterances \cite{Yamamoto}. 
However, only a minor improvement in performance could be obtained through this approach. 
We assume that this limitation occurred because it is difficult to compensate the missing phonetic information using already extracted utterance-level features \cite{uncertaintypropagation, Yamamoto, yang2017applying}.

Unlike most previous studies that compensate utterance-level features after they have been extracted, we propose a novel short utterance compensation system based on phonetic-level features. 
The proposed system extracts speaker embeddings directly from short utterances. 
The phonetic-level feature in this study is defined as an intermediate concept between frame-level and utterance-level features that effectively represents phonetic information, which covers approximately 130 ms. 
The duration of 130 ms is known to be appropriate for representing phonetic information based on conventional phonetic knowledge \cite{peterson1960duration, ordin2015acquisition}. 
Figure 1-(a) illustrates the concept of the phonetic-level features. 

To efficiently compensate the short utterances using phonetic information, we use the convolutional neural network long short-term memory (CNN-LSTM) architecture proposed by Jung \textit{et al.} with a few modifications \cite{jung2018avoiding}. 
This model directly extracts utterance-level embeddings from raw waveform, where the process can be divided into frame-level feature extraction and utterance-level feature aggregation. 
The CNN is used to conduct the former and LSTM is used for the latter. 
Here, we define the output of the last convolutional layer as the phonetic-feature. 
Using this model, teacher-student (TS) learning framework is conducted where cosine distance of speaker embeddings from long and short utterances are compared to efficiently compensate the short utterances using phonetic information. 
Resulting proposed system is an integrated short utterance compensation system that extracts speaker embeddings directly from short utterances of 2.05 s duration, text-independently (overall illustration in Figure 1-(b)). 

The remaining paper is organized as follows: 
Section 2 describes the speaker embedding system. 
Section 3 introduces the teacher-student learning framework. 
The proposed short utterance compensation system is discussed in Section 4. 
The experimental settings and result analysis are described in Section 5 and the study is concluded in Section 6.

\section{Speaker embedding model}
\label{sec:2}
Recent advances in deep neural networks (DNNs) have resulted in several successful speaker embedding systems that directly model raw waveforms \cite{jung2018avoiding, jung2018complete, muckenhirn2018towards}. 
These studies have shown that suitable pre-processing for speaker verification could be performed, yielding comparable or better results than conventional Mel-energy feature or spectrogram-based systems \cite{jung2018avoiding, ravanelli2018interpretable}. 
In this study, we use the raw waveform CNN-LSTM (RWCNN-LSTM) architecture proposed in \cite{jung2018avoiding} with the following two modifications: leaky rectified linear unit (LReLU) activation was used \cite{maas2013rectifier} instead of ReLU activation, and the long short-term memory layer was replaced with a GRU layer. 
Comparative experimental results show that these two modifications lead to an additional decrease of 10 \% in terms of equal error rates (EERs). 
Note that other DNN systems can also used for the short utterance compensation scheme proposed in Section 4. 

\begin{figure}[t!]
  \centering
  \centerline{\includegraphics[width=0.9\columnwidth]{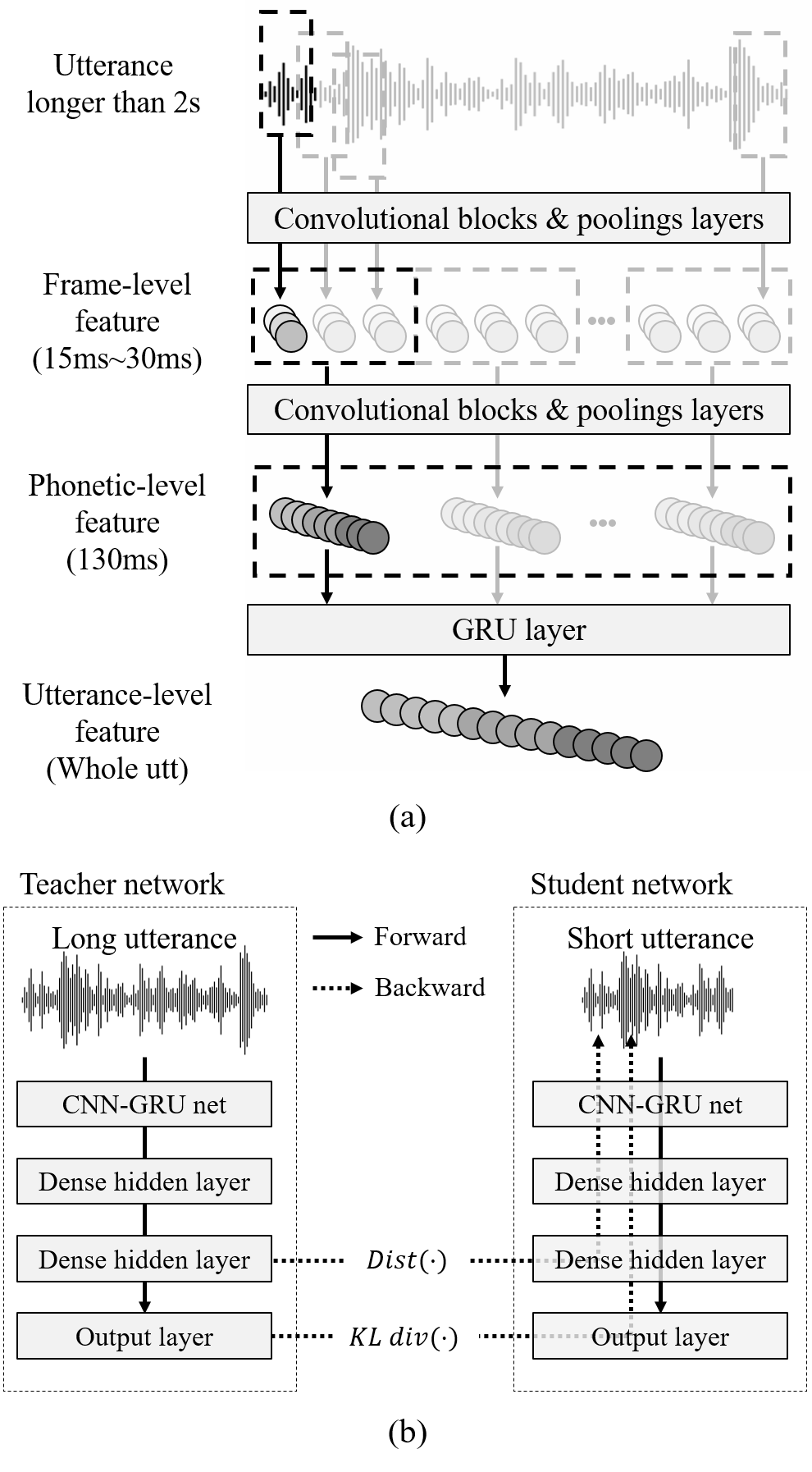}}
  \caption{(a) Conceptual illustration of various levels of features based on CNN-GRU network (b) Workflow of the proposed teacher-student learning-based short utterance compensation system.}
  \label{fig1}
\end{figure}

The RWCNN-GRU model comprises convolutional blocks followed by one GRU layer and two fully-connected layers. 
Each convolutional block has a residual connection \cite{Residual} and contains a pooling layer at its end, same with the convolutional block used in \cite{jung2018avoiding}. 
The last convolutional block transforms raw waveforms into phonetic-level features (addressed in Section 4) that represent segments of approximately 130 ms. 
The outputs of the last convolutional block are fed to the following GRU layer, which produces fixed low dimensional utterance-level representations. 
Then, the last fully-connected layer's LReLU activation is used as the speaker embedding. 
Speaker verification is performed by comparing the cosine distance between two speaker embeddings. 
In this research, both teacher and student DNNs have an identical architecture. 
However, the sequence length of the output of the last convolutional block (which can also be thought of the timestep of the GRU input) varies depending on the length of input utterances. 

\section{Teacher-student learning}
\label{sec:3}

Teacher-student (TS) learning uses two DNNs, teacher and student, in which the student DNN is trained using soft labels that the pre-trained teacher DNN provides. 
In this framework, after the teacher network is trained, the student network is trained to have an output distribution similar to that of the teacher network. 
This framework was first proposed for model compression and is also widely used for compensating far-field utterances \cite{li2014learning, li2018developing, kim2018bridgenets}. 
Adoption of the TS framework into short utterance compensation is novel in our knowledge. 
When TS learning is used for short utterance compensation, the Kullback–Leibler (KL) divergence objective function can be written as 
\begin{equation}
  \label{eqn:short utterance ts learning}
    \displaystyle KL_{loss} = - \sum\limits_j^J\sum\limits_i^I p_T(s_i | x_{j, l}) \log(p_S(s_i | x_{j,sh})),
\end{equation}
\noindent where $i$ and $j$ refer to the speaker and utterance indices, respectively; $x_l$ and $x_{sh}$ refer to the long and short crop of the same utterance, respectively; and $p_T(s_i|\cdot)$ and $p_S(s_i|\cdot)$ are the outputs of the teacher and student DNNs, respectively. 
The above equation shows that TS learning trains the student DNN's output layer distribution same as that of the teacher DNN despite being provided with short utterances.

\section{Proposed short utterance compensation system}
\label{sec:4.}

\begin{figure}[t!]
  \centering
  \centerline{\includegraphics[width=0.9\columnwidth]{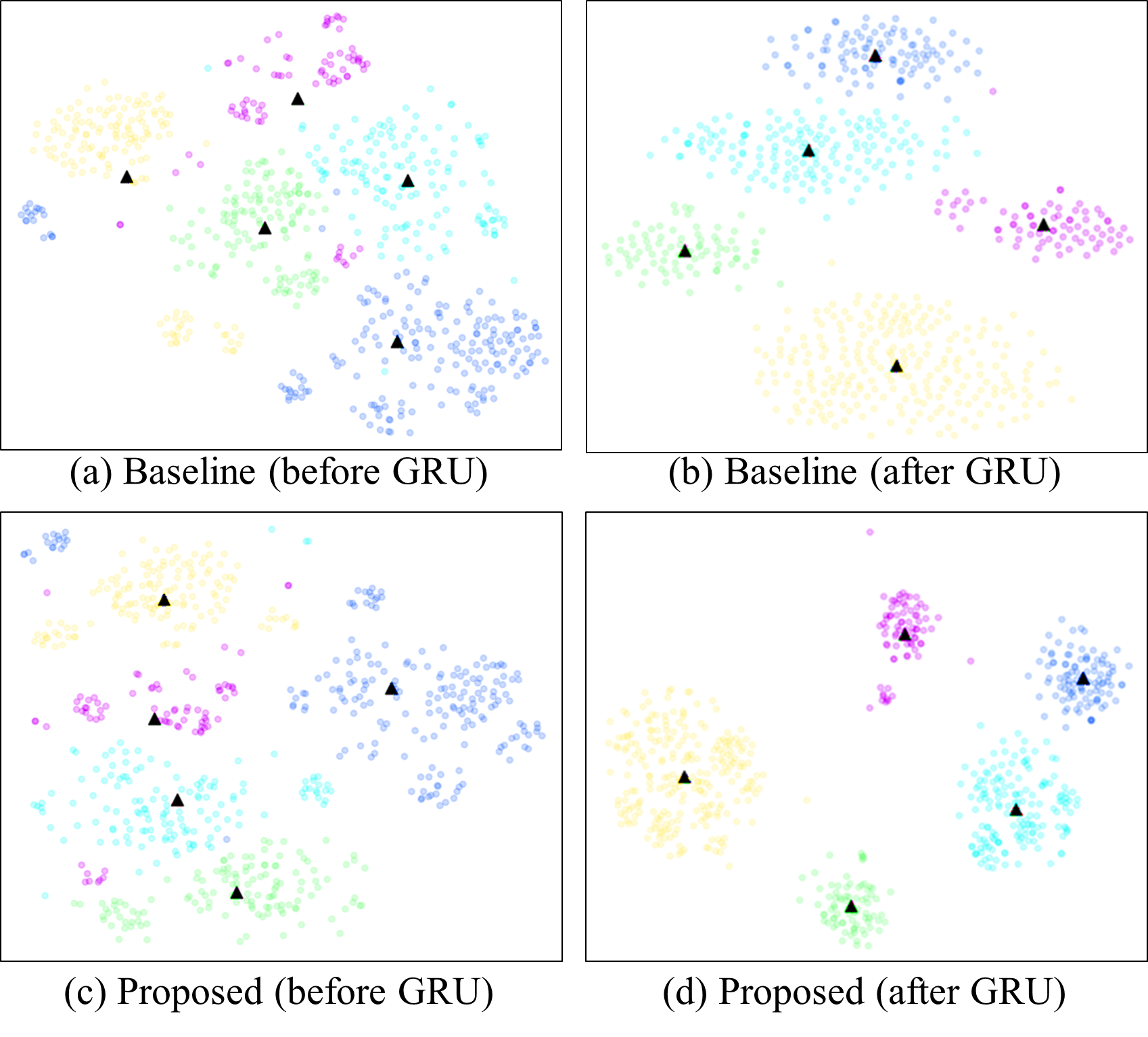}}
  \caption{Speaker embeddings visualized using the t-SNE algorithm \cite{maaten2008visualizing}. The five different colors represent five randomly selected speakers from the evaluation set. A triangle denotes the mean of the speaker embeddings extracted from long utterances.}
  \label{fig2}
\end{figure}

The main goal of the proposed system is to conduct the compensation throughout the network, rather than compensating via a separate system after extracting the speaker embedding. 
TS learning addressed in Section 3 is applied to short utterance compensation for this goal. 
To compensate more efficiently, we propose a direct compensation using the speaker embeddings and the output layer while conventional TS learning only uses the output layer. 
This is because the ultimate goal of a short utterance compensation system is to make the speaker embedding of a short utterance identical to that of a long utterance, comparison of speaker embeddings would be a more direct approach. 
% However, because the ultimate goal of a short utterance compensation system is to make the speaker embedding of a short utterance identical to that of a long utterance, we propose an objective function to compare speaker embeddings directly. 
The objective function of the proposed TS learning can be written as an extension of Equation 1, 
\begin{equation}
\begin{split}
  \label{eqn:proposed objective function}
    \displaystyle Loss = \sum\limits_j^J Dist(p_T(e | x_{j, l}), p_S(e | x_{j, sh})) 
    \\- \sum\limits_j^J\sum\limits_i^I p_T(s_i | x_{j, l}) \log(p_S(s_i | x_{j,sh})).
\end{split}
\end{equation}
Here $p_T(e|x_{j, l})$ and $p_S(e|x_{j,sh})$ denote the speaker embedding of the teacher and student DNNs, respectively, and $Dist(\cdot,\cdot)$ denotes the measure of the distance between two embeddings such as the cosine distance or mean squared error. 

The approach presented herein is notably different from existing short utterance compensation approaches owing to two aspects. 
The first is that short utterances are compensated throughout the entire DNN, mainly phonetic-level and GRU layer, rather than being compensated after extracting utterance-level features. 
Previous researches exploited an additional compensation system to transform speaker embeddings extracted from short utterances after utterance-level feature extraction. 
This is because the uncertainty caused by lacking phonetic information is observed in utterance-level features. 
However, compensating phonetic-level features appears to be a more direct solution, because uncertainty arises in the process of extracting utterance-level features from phonetic-level features. 
We argue that using the proposed approach, although the transformation is performed throughout the network, compensation is mainly conducted in the GRU layer where it tries to move the utterance-level features of the short utterances to the optimal position derived from the corresponding long utterance with abundant phonetic information. 
Plots in Figure 2 is used to reinforce this argument. 

Figure 2 demonstrates the speaker embeddings before (left column), and after (right column) the GRU layer of the baseline (w/o TS, upper row) and the proposed model (w TS, lower row). 
Because the embeddings are from the evaluation set, unseen data, we expect that cohesiveness of each speaker's embeddings directly demonstrates the discriminative power. 
By comparing (a), and (b), we can conclude that the GRU layer increases the discriminative power for each speaker in both baseline and the proposed system. 
However, Figure 2 shows that the compensation is mainly conducted in the GRU layer because the difference of cohesiveness between (b), (d) is more noticeable than that between (a), (c).

The second difference pertains to the adoption of the approaches of compensating short utterances and maintaining the discriminative power simultaneously using the proposed TS learning approach with the proposed objective function. 
In \cite{zhang2018vector}, Jiacen \textit{et al.} reported that when short utterance compensation is performed, the speaker embedding of the short utterance become close to that of a long utterance. 
However, even though the distance between short and long utterances became closer in terms of a distance measure, the discriminative power of the compensated embedding could not be ensured. 
Our experimental results also confirmed that solely reducing the distance between two embeddings of long and short utterances did improve the performance, although not considerably. 
Therefore, to maintain the discriminative power of the speaker embedding when short utterance compensation is performed, KL-divergence term is included in the proposed objective function. 
This results in the speaker embedding layers being compared using the cosine distance metric (compensation), while also using the conventional KL-divergence loss (discriminative power), which is novel. 
Superior results were obtained using both losses as the final objective function. 
Overall illustration of the proposed system is depicted in Figure 1-(b).

\section{Experiments}
\label{sec:4}
\subsection{Dataset}
\label{ssec:4.1.}
In all the experiments described herein, we used the \texttt{VoxCeleb1} dataset, which comprises approximately 330 hours of audio of 1,251 speakers, at a sampling rate of 16 kHz \cite{Voxceleb}. 
The dataset involves utterances with an average and minimum duration of 8.2 s and 4 s, respectively, in a text-independent scenario. 
Our evaluation trials and training / evaluation subset divisions follow the dataset’s guidelines. 
To evaluate the performance on the long and short utterances, utterances of the evaluation set were cropped into lengths of 3.59 s (59,049 samples) and 2.05 s (32,805 samples), both enrollment and test utterances. 
We took the center part of each utterance to compose evaluation sets.

\subsection{Experiment configurations}
\label{ssec:4.2.}
The systems were implemented using Keras, which is a python library with a Tensorflow backend \cite{keras, tensorflow, tensorflow2}. 
We used the RWCNN-LSTM system \cite{jung2018avoiding} with two modifications for both teacher and student DNN architectures. 
The teacher DNN inputs the raw waveform corresponding to 59,049 samples ($\approx$ 3.59 s). It involves one strided convolutional layer with stride size of 3 and six residual convolution blocks that do not reduce the length of the input sequence (the residual block is identical to that employed in \cite{jung2018avoiding}). 
After each residual convolution block, a max pooling layer with stride size of 3 is applied. 
The output shape of the last convolution block is (27, 512) where 27 is the sequence length and 512 is the number of kernels in the last convolutional layer. 
$27$ is derived from $59,049 / (3 \times 3^6)$ where 59,049 is the number of samples, $3$ is for strided convolution, and $3^6$ for six max pooling layers.
We note that the number of phonetic-level embeddings extracted using CNN is fixed to 27 in training phase for batch construction of utterances of 59,049 samples, but can vary at evaluation phase depending on the duration of each utterance (e.g. each 2,187 samples yield one phonetic-level embedding, utterance of 2.05 s duration will yield 15 phonetic-level embeddings). 
The GRU layer has 512 units and the two fully-connected layers have 1,024 nodes each. 
The multi-step training proposed in \cite{jung2018avoiding, E2E_Heesoo} is used for training the teacher DNN. 
The weights of the teacher DNN are frozen when the student DNN is trained. 

The student DNN is initialized using the weights of the teacher DNN as this process has been proved to ease the training in \cite{pang2018compression}. 
The architecture of the student DNN is identical to that of the teacher DNN except that the student DNN inputs raw waveform with 32,805 samples ($\approx$ 2.05 s), which means that the output shape of the last convolution block is (15, 512). 
When training the student DNN, two mini-batches where one comprises utterances of 59,049 samples and the other comprises 32,805 samples are respectively fed into teacher and student DNN. 
Then, cosine distance and KL-divergence are calculated using the last hidden layers and output layers of teacher and student DNN.  

The stochastic gradient descent with learning rate of 0.001 and momentum of 0.9 was used as the optimizer when training the teacher DNN. 
The same optimizer with a learning rate of 0.01 was used for training the student DNN.

\subsection{Results and analysis}
\label{ssec:4.3.}

The baseline performances are presented in Table 1. 
Using \texttt{VoxCeleb1} evaluation set without duration restriction which comprises approximately 3 s to 7 s duration, EER of 7.51 \% was obtained. 
EER increased by 46 \% , relatively, when the duration of the evaluation set was changed from 3.59 s to 2.05 s, which shows performance degradation owing to the short duration (8.72 \% to 12.8 \%). 
The research objective in this study is to compensate EER of 12.8 \% to 8.72 \%. 

Experimental result of training with short utterances at the first time, one of the well-known approaches for short utterance compensation, is shown in the lower row of Table 1. 
Performance did improve, but only minor improvement of 5 \% relative reduction in terms of EER was obtained. 
This result seems to have occurred because the duration of the short utterance considered herein is less than that used in other studies, in a text-independent scenario (configurations of 5 or 10 s are usually used) \cite{Yamamoto}. 

\begin{table}[h]
  \caption{Performance of the baseline systems with different durations. ``Full-length eval'' corresponds to the use of various length utterances without modification. The numbers represent EERs (\%).}
  \label{tab:tale1}
  \setlength{\tabcolsep}{11pt}

  \centering
  \begin{tabular}{c c c c}
    \Xhline{3\arrayrulewidth}
   
    \multirow{2}{*}{\textbf{System}} & \textbf{full-length} & \textbf{3.59 s} & \textbf{2.05 s}\\
    & \textbf{eval} & \textbf{eval} & \textbf{eval} \\
    \hline
    RWCNN-GRU   & \multirow{2}{*}{7.51} & \multirow{2}{*}{8.72} & \multirow{2}{*}{12.80}\\
    (3.59 s train)     \\
    RWCNN-GRU         & \multirow{2}{*}{-} &\multirow{2}{*}{-} & \multirow{2}{*}{12.08}\\
    (2.05 s train)      &                       & & \\
    \Xhline{3\arrayrulewidth}
  \end{tabular}
\end{table}
    
Table 2 presents the results of the proposed approaches. 
Conventional TS learning, which uses the output layer's KL-divergence loss, did not show noticeable improvement. 
The proposed method that directly compares the speaker embedding layers demonstrated a better performance (using only the `dist' term in Equation 2), with EER 10.98 \% and 10.8 \% for mean squared error and cosine distance as distance metrics, respectively. 
The best result could be achieved by using both the KL-divergence of the output layer and cosine distance of speaker embedding layer, which compensated more than 65 \% of the performance degradation due to shortened input utterance. 
We interpret that the reason for additional performance increase by comparing both output and speaker embedding can be found in Jiacen \textit{et al.}'s research \cite{zhang2018vector}. 
This research suggested that when compensating short utterances, the compensated feature can become similar to that of the long utterance in terms of the distance scale used for compensation (i.e. Euclidean), but this may not lead to increase in its discriminative power. 
In other words, in our interpretation, it means that usage of distance metric alone cannot consider the manifold structure of the speaker embedding space. 
Referring to this argument, comparing speaker embeddings can make the embedding of the student DNN equivalent to that of the teacher DNN, and the KL-divergence between output layers can help maintain its discriminative power. 

\begin{table}[h]
  \caption{Evaluation of various proposed systems using the modified 2.05 s evaluation set. ``Embedding'' and ``Output'' refer to layers to compare between the teacher and student networks. Values inside the bracket indicates the metric.}
  \label{tab:tale2}
  \setlength{\tabcolsep}{11pt}

  \centering
  \begin{tabular}{l c}
    \Xhline{3\arrayrulewidth}
      
    \textbf{Systems} & \textbf{EER} (\%)\\
    \hline
    Output (KL-Div) (Original TS) & 12.46\\
    Embedding (MSE) & 10.98\\
    Embedding (Cos Sim) & 10.80\\
    Embedding (Cos Sim)+Output (KL-Div) & 10.08\\
      
    \Xhline{3\arrayrulewidth}
  \end{tabular}
\end{table}

Additionally, Table 3 shows the evaluation using varying duration utterances. 
It is not realistic to fix the duration of utterance in real world applications, which makes less duration variant systems necessary. 
To verify how invariant the proposed system is towards varying duration short utterances, experiments with different range of duration have been conducted. 
Results demonstrate that EER of both baseline system (w/o TS, upper) and the proposed system (w TS, lower) are not much variant to the duration of input utterance. 
We note that the performance degradation as range widens is considered as the effect of inclusion of shorter utterances (e.g. duration starts from 1.55 s in ``1.55$\pm$0.5 s''. 

  \begin{table}[h]
\caption{Evaluation with varying utterance duration. Performance degradation as range widens is due to shorter utterances (e.g. duration starts from 1.55 s in ``2.05$\pm$0.5 s'').}
\label{tab:table4}
\setlength{\tabcolsep}{11pt}

\centering
\begin{tabular}{c c c c}
    \Xhline{3\arrayrulewidth}
    \multirow{2}{*}{\textbf{System}} & \multirow{2}{*}{\textbf{2.05}} & \textbf{2.05} & \textbf{2.05}\\
    &  & \textbf{$\pm$0.1 s} & \textbf{$\pm$0.5 s}\\
    \hline
    RWCNN-GRU   & \multirow{2}{*}{12.80} & \multirow{2}{*}{12.96} & \multirow{2}{*}{13.28}\\
    (w/o TS)     \\
    RWCNN-GRU   & \multirow{2}{*}{10.08} & \multirow{2}{*}{10.29} & \multirow{2}{*}{10.40}\\
    (w TS)     \\
    \Xhline{3\arrayrulewidth}
\end{tabular}
\end{table}

\section{Discussion and future work}
\label{sec:5}
In this paper, we proposed a text-independent short utterance speaker verification system that works on utterances with durations of 2.05 s or less. 
The proposed system does not transform the utterance-level feature from the short utterance as in conventional approaches. 
Rather, it directly extracts the compensated speaker embeddings from short utterances by compensating throughout the network, focusing on phonetic-level compensation. 
This is because we expected that the main key for compensating short utterances corresponds to the phonetic information, whose absence leads to the uncertainty of the utterance-level features. 
To process phonetic information, phonetic-level features that represent segments of 130 ms were extracted using CNN, and then transformed to the utterance-level features using a GRU layer. 
The effectiveness of the defined phonetic-level features was verified by the performance improvement of the speaker verification system using short utterance compensation and an illustration of the cohesiveness of speaker embeddings from the evaluation set. 
An objective function is also proposed to conduct a more effective compensation by considering distance of long and short utterance and concurrently maintain the discriminative power of speaker embeddings. 

In the future, we will analyze the information included in phonetic-level features and construct phonetic-level features using speech recognition systems. 
Additionally, because the proposed compensation scheme can be applied to any DNN-based speaker embedding extraction schemes, such as Mel-energy feature-based x-vector or other spectrogram-based systems, we plan to apply the proposed scheme into other systems.

\vfill\pagebreak

\bibliographystyle{IEEEtran}
\bibliography{mybib}

\end{document}